# Strong asymmetry of forward scattering effect from dielectric cubic nanoantenna in lossless media

Pavel D. Terekhov,[†‡¶§*] Hadi K. Shamkhi,[‡*] Egor Gurvitz,[‡] Kseniia Baryshnikova,[‡] Andrey B. Evlyukhin,[‡‖] Alexander S. Shalin,[‡] Alina Karabchevsky[†¶§Δ]

*†Electrooptics and Photonics Engineering Department, Ben-Gurion University, Beer-Sheva 8410501, Israel*

*‡ ITMO University, 49 Kronverksky Ave., 197101, St. Petersburg, Russia*

*¶ Ilse Katz Institute for Nanoscale Science & Technology, Ben-Gurion University, Beer-Sheva 8410501, Israel*

*§ Center for Quantum Information Science and Technology, Ben-Gurion University, Beer-Sheva 8410501, Israel*

*‖ Moscow Institute of Physics and Technology, 9 Institutsky Lane, Dolgoprudny 141700, Russia*

*Authors contributed equally

[Δ]alinak@bgu.ac.il

## Abstract

Dielectric photonics platform provides unique possibilities to control light scattering via utilizing high-index dielectric nanoantennas with peculiar optical signatures. Despite the intensively growing field of all-dielectric nanophotonics, it is still unclear how surrounding media affect scattering properties of a nanoantenna with complex multipole response. Here we report on light scattering by a silicon cubic nanoparticles embedded in lossless media, supporting optical resonant response and. We show that significant changes in the scattering process are governed by the electro-magnetic multipole resonances which experience spectral red-shift and broadening over the whole visible and near-infrared spectra as the indices of media increase. Most interestingly that the considered nanoantenna exhibits the broadband forward scattering in the visible and near-infrared spectral ranges due to the Kerker-effect in high-index media. The revealed effect of broadband forward scattering is essential for highly demanding applications in which the influence of the media is crucial such as health-care: sensing, treatment efficiency monitoring, and diagnostics. In addition, the insights from this study are expected to pave the way towards engineering the nanophotonic systems including but not limited to Huygens-metasurfaces in media within a single framework.

## Introduction

Tuning the scattering resonances of plasmonic or dielectric nanoparticles has enabled scientists and engineers to localize light at nanoscale and enhance peculiar optical phenomena over the last decade[1,2]. Specifically, changing the shape and dimensions of high-index dielectric nanoparticles leads to the variation of their optical properties opening a door for a plethora of interesting phenomena [2–7]. Nanoparticles scatter light and support the excitation of the geometrical (Mie) electric and magnetic multipolar resonances[2–8] which in turn enhance light-matter interaction. Despite the fact that the influence of nanoparticles size, geometry, and material on their optical properties is being extensively explored[3,9,10] for cloaking[11] and spectroscopy[12], the influence of media in which the nanoparticles are embedded is still obscured. Here, therefore, we bridge this gap in knowledge by exploring the optical properties of high-index nanoparticles depending on the optical properties of a surrounding medium. One of the auxiliary tools for understanding the interaction of light with nanoparticles is multipole decomposition[13].

The role of multipole contributions to scattering effect is of high importance for a variety of applications, including but not limited to nanoantennas[14–18], sensors[19–21], solar cells[22], multi-functional metasurfaces[23] and cloaking devices[11,24]. Dielectric nanoparticles could be used for a drug delivery, and as probes for an electromagnetic diagnosis in emerging medical applications[25–27]. The particles' resonant behavior in dielectric media like, e.g., the human body environment is another very interesting question[28].

Forward scattering phenomenon has been a subject of extensive studies both theoretically and experimentally[29–32]. Most of such studies are based on engineering the multipolar resonances to meet the requirement of the well-known Kerker effect[29]. In a core-shell nanoparticle, for instance, the dielectric shell shifts the electric dipole resonance of the metal core so both core and shell resonances overlap and contribute to the forward scattering due to constructive interference[31]. A high-index substrate on another hand provides a leakage medium and results in a broadening of the multipoles resonances, therefore supporting broadband forward scattering and allowing a design of efficient dielectric Huygens' metasurfaces[30,32].

In this work, we show that reducing the refractive index contrast of the particle-medium system along with the silicon nanocube optical resonances leads to important spectral features such as shifting and broadening of the multipole resonances. The proposed structure supports overlapping of several multipole resonances leading to the strong asymmetric broadband forward scattering. Insights from this study are expected to pave the way towards engineering the nanophotonic systems in media within a single framework.

## Theoretical Background

In this paper we consider the optical system including the single silicon nanocube suspended in a lossless medium. Figure 1 shows the schematics of the studied system which is a nanocube with edge of $H = 250\ nm$, made of polycrystalline silicon[33] and suspended in media with the indexes $n$ of $1 \leq n \leq 2$ having increment of $dn = 0.2$. Note that the wavelengths values throughout the article are considered for the incident light wave in air.

To analyze the considered system, we use the semi-analytical multipole decomposition approach reported in Refs.[34]. This approach is based on the multipole expansions frequently used in electrodynamic theory.[34–36] Here we use the expressions of the dynamic multipole moments up to the magnetic quadrupoles as defined in Ref.[36] Note that for scatterers located in a medium the wavenumber, introducing in the expressions for multipoles,[36] should be taken in this medium. In addition, we consider the quasistatic electric octupole moment as derived in Ref [34]. We validated that the considered set of multipole moments is sufficient for the proper description of the system we study. In contradiction, the quasistatic approximation is not suitable, because long-wavelength approximation cannot be applied as we show in the Supplementary material. Figure S1 of the Supplementary material shows scattering cross-sections which we calculated as sum of multipoles contributions using two different approaches 1) dynamic approximation and 2) quasistatic approximation, and their comparison with scattering cross-section calculated as ratio of corresponding power flows for media with n = 1 (left) and n = 2 (right). The full electric field in the system is numerically calculated by the finite elements method (FEM) implemented in the COMSOL Multiphysics commercial package[37,38]. The calculated electric field is used for obtaining of the multipole moments and their contributions in the scattering cross sections.[34–36]

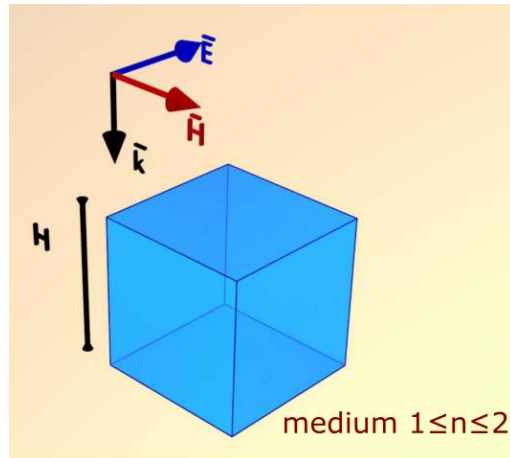

Figure 1: The schematics of the single particle suspended in a medium.

## Asymmetry parameter

The main goal of our work is studying of the directional scattering effect in lossless media. Such effects like Kerker-effect are of interest due to asymmetric light scattering in the forward and backward directions. For the convenient description of such effects, we introduce the asymmetry parameter. The asymmetry parameter is the efficiency factor that quantifies the directivity of light scattering. To derive and calculate the asymmetry parameter for an arbitrarily shaped particle, we apply the Cartesian multipole decomposition method for the scattered field. Then, the scattered electric field $\mathbf{E}_{\mathrm{sca}}$ by a particle in a homogenous host medium is the superposition of the induced multipole moments contributions:[34]

$$\mathbf{E}_{\mathrm{sca}}(\mathbf{r}) \cong \frac{e^{i\,k_d r}}{r}\frac{k_0^2}{4\pi\varepsilon_0}\bigg(\big[\mathbf{n}\times[\mathbf{D}\times\mathbf{n}]\big] + \frac{1}{v_d}[\mathbf{m}\times\mathbf{n}] + \frac{ik_d}{6}\big[\mathbf{n}\times\big[\mathbf{n}\times(\hat{Q}\cdot\mathbf{n})\big]\big]$$
$$+ \frac{ik_d}{2v_d}\big[\mathbf{n}\times(\hat{M}\cdot\mathbf{n})\big] + \frac{k_d^2}{6}\big[\mathbf{n}\times\big[\mathbf{n}\times(\hat{O}\cdot\mathbf{n}\cdot\mathbf{n})\big]\big]\bigg),$$

$$(1)$$

where the unit vector $\mathbf{n}$ is in the direction of scattering vector $\mathbf{r}$ , $\varepsilon_d$ is the relative dielectric permittivity of the surrounding medium, $\varepsilon_0$ is the vacuum electric permittivity, $v_d = c/\sqrt{\varepsilon_d}$ is the light speed in the surrounding medium and $c$ is the light speed in vacuum; $k_0$ and $k_d$ are the wavenumbers in vacuum and in the surrounding medium respectively; $\mathbf{m}$ is the magnetic dipole moment (MD); $\mathbf{D}$ is the total electric dipole moment (TED); $\hat{Q}$ , $\hat{M}$ and $\hat{O}$ are the electric quadrupole moment tensor (EQ), the magnetic quadrupole moment tensor (MQ) and the tensor of electric octupole moment (OCT), respectively. Note that these tensors are symmetric and traceless and in tensor notation e.g. $\hat{Q}$ is equal to $Q_{\alpha\beta}$, where subscript indices denote components (e.g $\alpha = \{x, y, z\}$) [34]. The scattering cross-section can be presented as (see [34] for details):

$$C_{sca} \cong \frac{k_0^4}{6\pi\varepsilon_0^2|\mathbf{E}_{\text{inc}}|^2}|\mathbf{D}|^2 + \frac{k_0^4\varepsilon_d\mu_0}{6\pi\varepsilon_0|\mathbf{E}_{\text{inc}}|^2}|\mathbf{m}|^2 + \frac{k_0^6\varepsilon_d}{720\pi\varepsilon_0^2|\mathbf{E}_{\text{inc}}|^2}|\hat{Q}|^2$$

$$+ \frac{k_0^6\varepsilon_d^2}{80\pi\varepsilon_0|\mathbf{E}_{\text{inc}}|^2}|\hat{M}|^2 + \frac{k_0^8\varepsilon_d^2}{1890\pi\varepsilon_0^2|\mathbf{E}_{\text{inc}}|^2}|\hat{O}|^2,$$

(2)

where $\mathbf{E}_{\text{inc}}$ is the electric field amplitude of the incident light wave. The total scattering cross-section is obtained through the integration of the Pointing vector over a closed surface in the far-field zone and the normalization to the incident field intensity[34]. The asymmetry parameter is the ratio of the cosine-weighted scattering cross section over the total scattering cross section $C_{sca}$ and can be calculated as

$$g = \frac{1}{|\mathbf{E}_{\text{inc}}|^2 C_{sca}} \int |\mathbf{E}_{\text{sca}}|^2 \ r^2 \cos\theta \ d\Omega$$

(3)

after the integrating over the solid angle $d\Omega = \sin\theta \ d\theta d\phi$ , we find

$$g \cong \frac{1}{|\mathbf{E}_{\text{inc}}|^2 C_{sca}} \frac{k_0^4}{360\pi\varepsilon_0^2 v_d^2} \Big[ 60 v_d \Re\{D_x m_y^* - D_y^* m_x\} - 6k_d v_d^2 \Im\{D_\alpha Q_{\alpha z}^*\}$$

$$- 18k_d \Im\{m_\alpha M_{\alpha z}^*\} - k_d^2 v_d \Re\{Q_{y\alpha}^* M_{x\alpha} - Q_{x\alpha}^* M_{y\alpha}\}$$

$$- \frac{24}{315} k_d^3 v_d^2 \Im\{Q_{\beta\beta} O_{\alpha\alpha z}^* + 2Q_{\beta z} O_{\beta\alpha\alpha}^* - 5Q_{\alpha\beta} O_{\alpha\beta z}^*\} \Big],$$

(4)

where we used the properties of the involved multipole tensors - the symmetry and traceless. Physically, $I_{inc}(1-g)C_{sca}$ is the net rate of momentum transferring to a particle in the direction of the propagation where $I_{inc}$ is the irradiance of the incident light beam. Upon inspection of the asymmetry parameter formula presented above, we found that all the multipoles' cross-terms scattering light symmetrically along the polar angle $\theta = 90^O$ have vanished. Therefore, the asymmetry parameter is a measure of light directivity; meaning that, its zero value corresponds to the equal scattering in both half spaces. Predominating scattering to the lower half space (forward) and the upper half space (backward) results in positive and negative values, respectively. Furthermore, the aforementioned Kerker effect for dipoles and Kerker-like effects for high-order multipoles can be understood with the asymmetry parameter equations[39,40]. If we consider the case of $E_x$ polarized incident light, then Kerker condition of ED and MD fulfills when the ratio $v_d D_x / m_y = 1$ with $g = 0.5$. Kerker-like condition of TED and EQ interplay fulfills at $6i D_x / k_d Q_{xz} = 1$ while for the MD and MQ interplay it fulfills when $2i m_y / k_d M_{yz} = 1$. Consequently, the EQ and MQ scatter in forward directions when $v_d Q_{xz} / 3M_{yz} = 1$ and finally for EQ and EO coherence, the condition is $5Q_{xz} / 4i k_d O_{zzx} = 1$. Generally, the synchronous multipole couplings enhance the field directivity, and play a key role for the suppression of the backscattering. Here, the total electric fields and the corresponding induced polarization current in the scatterers are calculated numerically using COMSOL Multiphysics. Using the calculated polarization current, the multipole moments and their contributions to scattering cross-sections and the asymmetry parameter are obtained by the numerical integration.

In the next section we analyze in detail the results obtained for the different surrounding media and the multipoles behavior in the system.

## Results and discussion

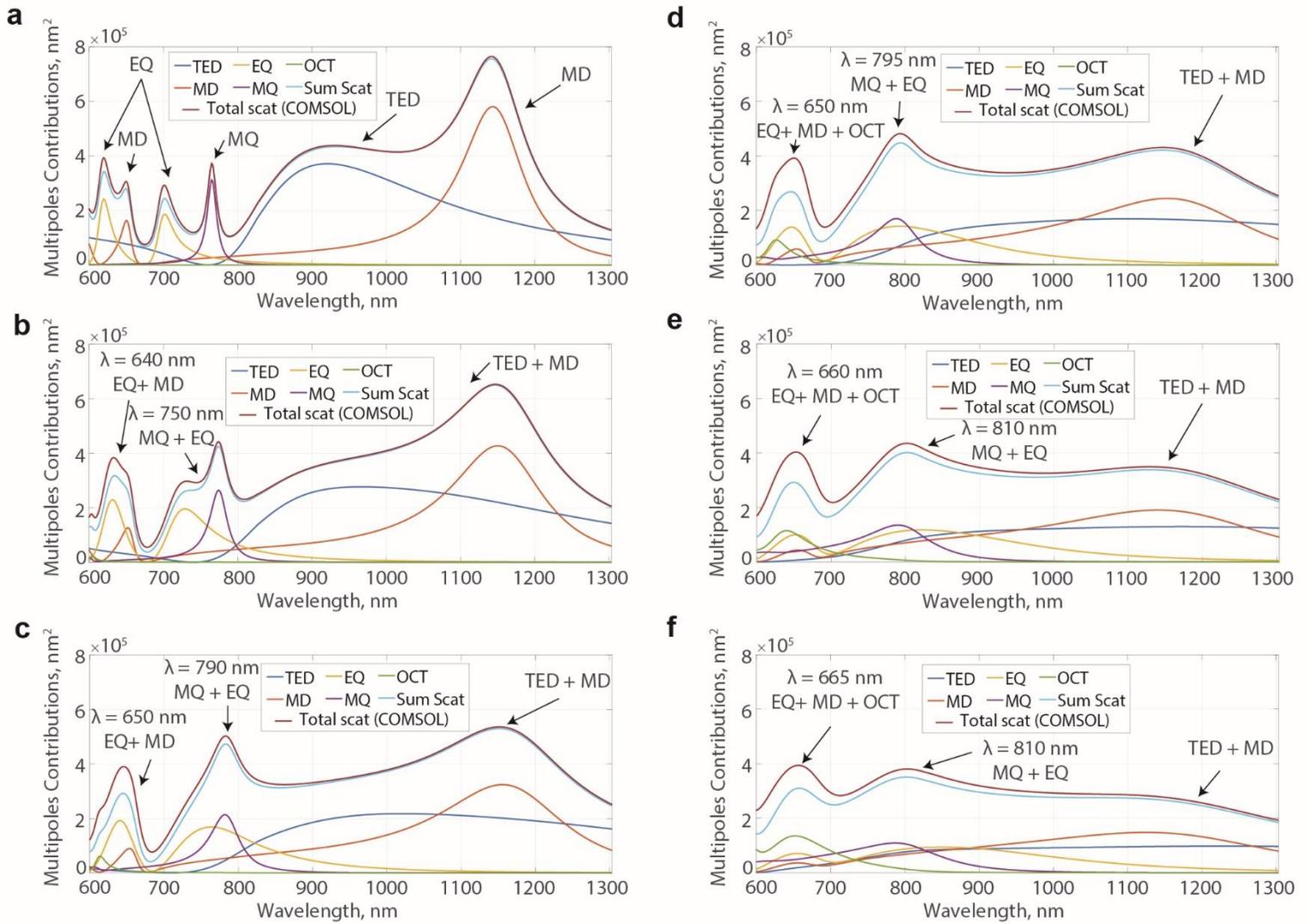

Figure 2: Total scattering cross-sections and the multipoles contributions calculated for the silicon nanocube embedded in different media. The refractive indexes of the surrounding media are (a) $n = 1$ (b) $n = 1.2$ (c) $n = 1.4$ (d) $n = 1.6$ (e) $n = 1.8$ (f) $n = 2$. "Sum Scat" states for the scattering cross-section as the sum of the multipole contributions; "Total scat (COMSOL)" states for the total scattering cross sections calculated directly in COMSOL. Black arrows mark the resonant areas and describe the dominant multipole contributions to the resonant scattering.

**Multipoles spectra evolution**

To analyze the optical properties of the considered nanoparticle in different media, we draw the multipole decomposition of the scattering cross-section spectra for the six different cases of $1 \leq n \leq 2$ (Fig. 2). To prove that our multipole decomposition approach can be applied, we compare the scattering cross-sections obtained as the sum of multipole contributions (Sum Scat) and the scattering cross-section obtained with the direct calculation (Total scat (COMSOL)) for every case. As discussed in the Supplementary material, this comparison shows the good coincidence between the calculation methods when we use the dynamic multipole moments.

We start our analysis with the case of the particle embedded in air. The case of $n$ = 1 (Fig. 2a) has been considered in details in[3] for the similar nanoantenna. One can note the pronounced resonant peaks in the scattering cross-section spectrum; as can be noted from Fig. 2, these peaks are associated with the resonant excitations of the total electric dipole (TED) moment, the magnetic dipole (MD) moment, the electric quadrupole (EQ) moment and the magnetic quadrupole (MQ) moment. Fig. 2b shows that starting from $n = 1.2$ the scattering cross-section resonant peaks start to merge with each other. At $\lambda = 640 \ nm$, the scattering cross-section peak is now due to the interaction between MD and EQ moments forming the one wider scattering peak. Similarly, the EQ and MQ moments resonant peaks begin to merge into the single peak around $\lambda = 750 \ nm$, but for $n = 1.2$ they still can be distinguished. At last, the TED and MD resonances in the near-infrared spectral range experience some broadening and become a bit smoother. However, the spectra for $n$ = 1.2 is still close to the case of air medium.

For $n = 1.4$ (Fig. 2c), the MQ and EQ resonant peaks merge to the single scattering cross-section peak at $\lambda = 790 \ nm$. At the same time, the TED and MD resonances in the near-infrared spectral range continue their broadening; the TED resonant excitation no longer leads to a separate scattering cross-section peak. Finally, the EQ and MQ resonances continue to merge and provide the scattering cross-section peak together at $\lambda = 650\ nm$. In the case of $n = 1.4$ only three scattering peaks can be distinguished.

Next, we consider $n = 1.6$ (Fig. 2d). The EQ moment still experiences the bigger red shift in comparison with the magnetic multipole moments MQ and MD. For $n = 1.6$, one can note the resonant excitation of electric octupole moment (OCT) at $\lambda = 620\ nm$. We also expect the excitation of higher-order multipole moments in this wavelength region, but the moments taken into account are enough for a proper qualitative description of the system. It is worth noting that OCT contributes to the scattering peak formed by EQ and MD at $\lambda = 650\ nm$. At the region of longer wavelengths, the spectral positions of MQ and EQ resonances match each other at $\lambda = 795\ nm$ and the resulting scattering cross-section peak is well pronounced in Fig. 2d. The MD-induced scattering cross-section peak can be still noted at $\lambda = 1160$ nm. Such intermediate case still shows three scattering cross-section peaks however slightly broadened.

The final step of the study is presented in Fig. 2e,f for $n = 1.8$ and $n = 2$. In the wavelength range $600 \leq \lambda \leq 700\ nm$, the resonant excitation of OCT provides the incremental contribution to the scattering cross-section. This contribution, together with the EQ and MD resonant excitations and the non-resonant MQ contribution, forms the scattering cross-section peak at the considered region. The EQ moment resonance continues its shifting to the red zone with respect to the MQ resonant area; hence, their joint scattering cross-section peak becomes smoother ($\lambda = 810\ nm$). The MD-induced scattering peak at $\lambda \approx 1170\ nm$ is almost smoothed out for $n = 2$. The multipole analysis performed in this way is very important to understand the origins of scattering process. Obtained insights are necessary to study broadband forward scattering enchantment and explain it in detail.

To provide the better visibility of the multipole evolution in Fig. 3, and to improve the consequent analysis of the scattering asymmetry we show the evolution of each multipole contributions as $n$ changes. The evolution of the dipole moment contributions is shown in Fig. 3a,b for the TED and MD correspondingly; the evolution of the quadrupole moment contributions is shown in Fig. 3c,d for the EQ and MQ correspondingly. Fig. 3a clearly shows that the TED resonant area becomes broader, but the peak value decreases as $n$ rises. Moreover, the TED resonant region experiences a relatively strong redshift as $n$ rises. Similarly, both MD resonances at $\lambda \approx 660$ nm and $\lambda \approx 1150$ $nm$ (Fig. 3b) experiences broadening, but the maximum contribution decreases as $n$ rises. In addition, both resonant areas experience the little redshift.

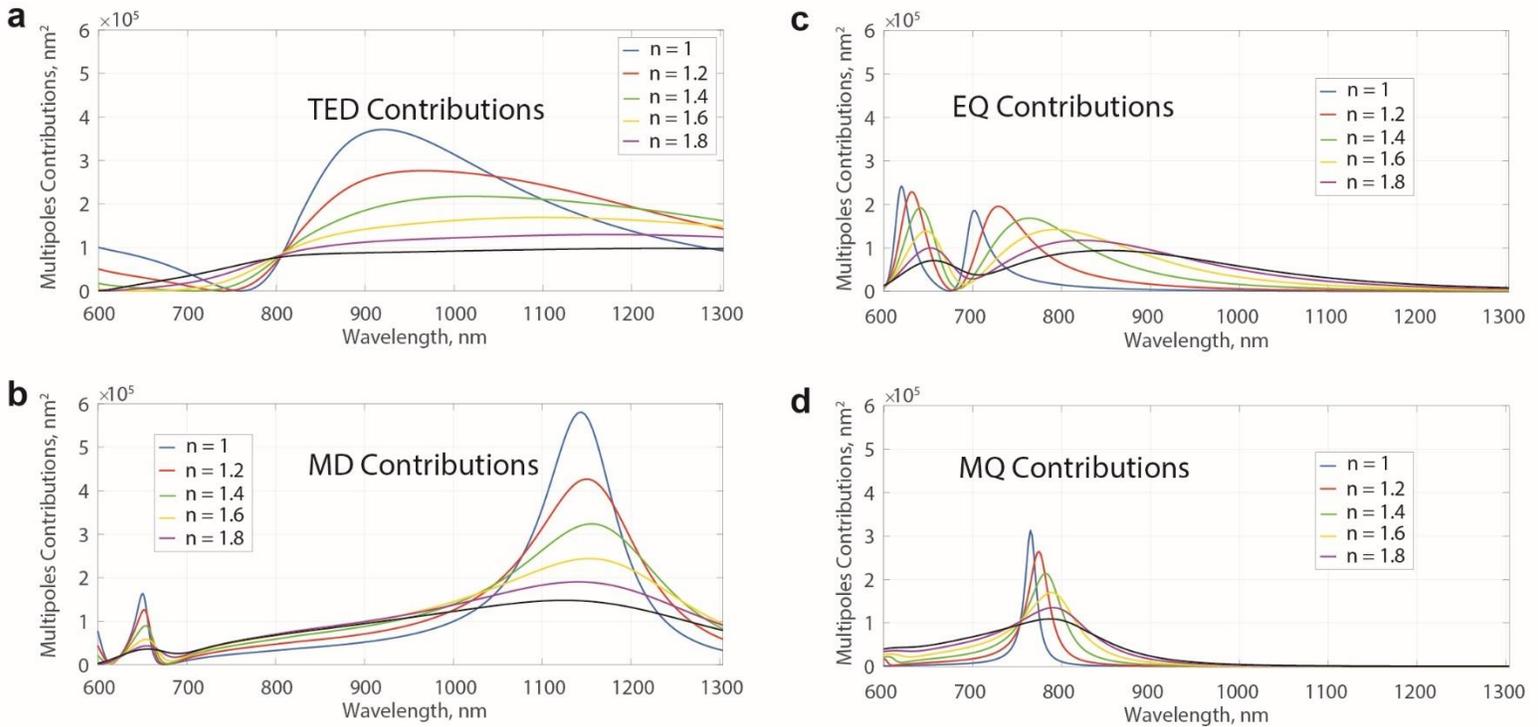

Figure 3: The multipole evolution of (a) the TED moment (b) the MD moment (c) the EQ moment (d) the MQ moment contribution to the scattering cross-section as the refractive index of surrounding medium rises.

Interestingly, the quadrupole resonances (Fig. 3c, d) behave slightly differently. As shown in Fig3c, both EQ resonances experience the redshift; it is especially strong for the resonant excitation at longer wavelengths, where the EQ resonant peak shifts from $\lambda = 700\ nm$ for $n = 1$ to $\lambda = 880\ nm$ for $n = 2$. The maximum value of the contribution to the scattering cross-section for this resonance decreases slower in comparison with TED moment as $n$ rises. Both EQ resonant regions experience some broadening. Further, the MQ moment has the only one resonant area for the considered system (Fig. 3d). This resonance experiences the noticeable redshift, but it is weak in comparison with the redshift of the EQ moment. The MQ resonance shows the broadening as $n$ rises, but the maximum value of the MQ contribution decreases. However, this decrease is slower in comparison with MD moment. Thus, we can associate the behavior of certain multipole moments with a change in refractive index of the media.

Such comparative analysis allows the study of the evolution of each multipole contribution to the scattering cross-section and to compare their behavior with each other. First, we prove that electric multipole resonances experience a stronger redshift than their magnetic counterparts as $n$ increases. Next, we show that the relative contributions of quadrupole moments to the scattering cross-section get stronger with *n*. These insights can be used to design specific combinations of the multipole moments for tuning the direction of the light scattering.

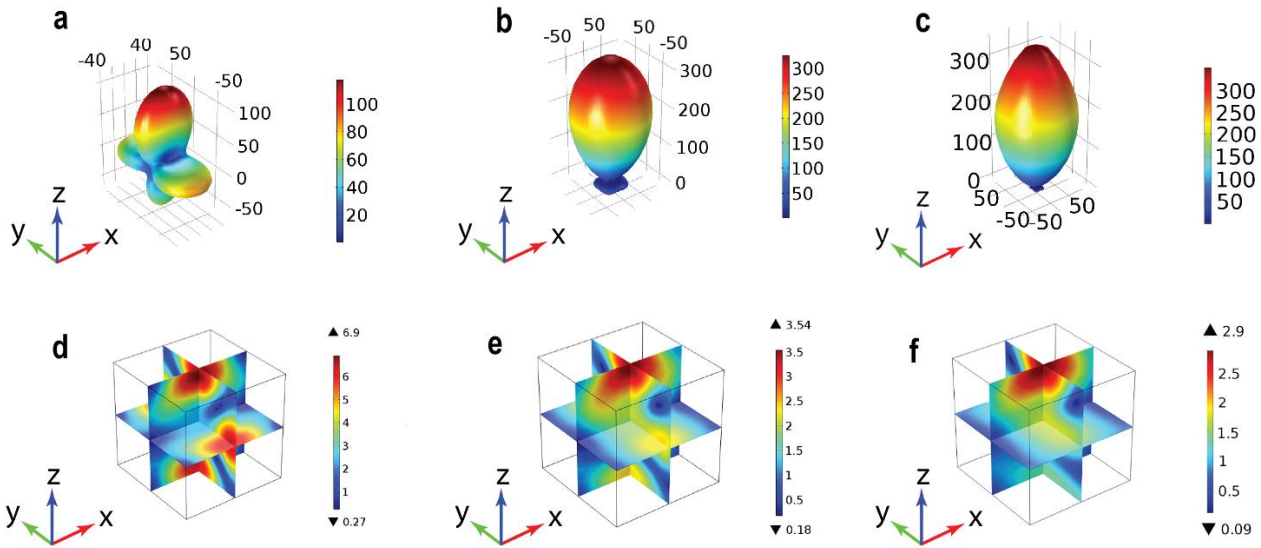

Figure 4: The radiation patterns and the electric field distribution inside the particle for (a, d) $n = 1, \lambda = 765\ nm$ (b, e) $n = 1.6, \lambda = 789\ nm$ (c, f) $n = 2, \lambda = 789\ n$.

We note that the spectral contributions of the multipole moments significantly change with *n*. In addition, the separated peaks merge, hence, the scattering peaks for higher *n* smooth out and broaden, which, as we will show in the next section, leads to the significant amplification of the forward scattering in a broad wavelength range. This enables us to design novel types of nanoantennas and antireflective coatings to apply in high-index surroundings and provides a room of opportunities for the design of new Huygens-type metasurfaces[30].

## Far-field scattering and electric field distribution inside the particle

In this section, we discuss the physical limitations of the forward scattering enhancement as refractive index contrast between the scatterer and the host media is reduced. The optical theorem states that the total energy extinct by a single nanoparticle is in direct relation to the scattering amplitude in the forward direction.[41]   Based on this definition we underline the energy conservation and physical limitations with the following explanations. The wave impedance of a medium is ($Z_0/n$) where $Z_0$ is the wave impedance in free space and hence high refractive index surrounding medium provide better coupling for the optical modes leaking out of the particle. Consequently, as the index contrast between the scatterer and the host media reduces, the inner reflection from the scatterer-medium interface decreases according to the Fresnel coefficients.[13] As a result, the induced multipole moments broadened and become less confined inside the nanoparticle. Fig.4 illustrates this effect by plotting the evolution of the field distribution both inside the nanocube and in the far-field region. The upper row (a,b,c) in Fig. 4 shows the evolution of the far-field scattering pattern at the wavelength corresponding to the the first MQ peaks in spectrum for the cases n=1, 1.6 and 2; in turn, the lower row (d,e,f) presents the field distribution inside the particle for the same cases. By selecting these points, we show the smooth transformation of the far-field radiation pattern. In the case of air medium MQ resonant contribution leads to the simultaneous scattering in four directions (see Fig. 2); however, the radiation pattern transforms as $n$ rises and exhibits the well-pronounced forward scattering effect with $n$ = 2. Here we analyse the evolution of the radiation pattern to emphasize the huge difference between two extreme cases.

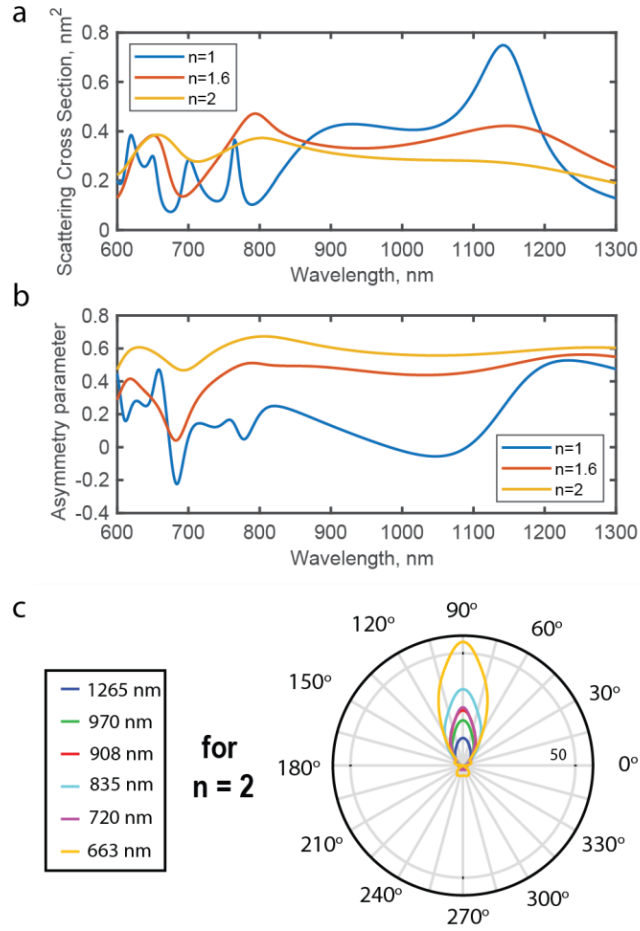

Figure 5: The scattering cross-section (a) and the asymmetry parameter (b) for $250\,nm$ silicon cube embeded in a medium with $n = 1, 1.6, 2$. (c) The 2D radiation patterns in the medium with $n = 2$ for various wavelengths $\lambda$ as indicated in the legend. The radiation patterns are presented in the plane of the electric field polarization of the incident wave.

In Fig. 5a and b, we present the scattering cross-sections (Eq. 2) for the three cases of $n = 1$, $n = 1.6$ and $n = 2$ along with the asymmetry parameter (Eq. 4). As expected, the multipole broadening afore-explained in Fig. 3 causes the synchronous overlapping of the multipoles of different orders and leads the asymmetry parameter to be dramatically broadened and enhanced as the surrounding medium refractive index increases. Balanced dipole moments fulfilling the Kerker's condition provide the asymmetry parameter $g = 0.5$; however, when the quadrupole contributions are significant, the asymmetry parameter becomes $g > 0.6$, indicating strong enhancement of the forward directivity. To clarify this point, in Fig. 5c we show the 2D radiation patterns for 6 different wavelengths over the spectrum for $n = 2$ and a strong asymmetry remains for all spectral points. Thereby, the forward scattering amplification in the broad range is the direct consequence of the high refractive index of the surrounding medium; the similar nanoantenna does not show such properties in case of $n = 1$.[3] Importantly, we note that the scattering efficiency for the cubical particle is also enhanced over the short wavelength optical range where the dipole, quadrupole and higher order multipoles contribute. However, for the long wavelength region, the scattering decreases since the higher order multipoles contribution is not enough to compensate the leaked dipoles. This result can be widely implemented to develop nanoantennas and other optical devices for applications in non-air media like liquids and dielectric materials. In addition, the broadening of the multipole moments especially the TED is of an importance for sensing applications. The weakening of the field inside the particle leads the multipole moments to be highly sensitive and exhibit redshift when the surrounding medium refractive index changes.

## Conclusion

To conclude, we reported on the broadband forward scattering effect due to the evolution of multipole moments for silicon cubical nanoparticles embedded in lossless media. We found that the electric multipole moments (TED, EQ) experience stronger redshift compared to the magnetic multipole moments (MD, MQ) as the index of medium increases. In addition, the separate scattering cross-section peaks transform to smoother merged peaks as the index of medium increases.

We discovered that with increasing the index of surrounding media, the broadband Kerker effect appears. We confirmed this effect through deriving the asymmetry factor for Cartesian multipoles system. We showed that the scattering intensity of the silicon nanoparticles is highly sensitive to the surrounding medium refractive index. Therefore, the studied system can work as an efficient sensor and is expected to be superior to the wavelength-shift depending plasmonic sensors. Also, the silicon nanocube as a building block in high-index media is expected to be applicable for the design of Huygens metasurfaces and transparent devices operating in the broadband spectral range.

## Acknowledgement

This work has been supported by the Israel Innovation Authority-Kamin Program, Grant. No. 62045. A.S.S acknowledges the support of the Russian Fund for Basic Research within the projects 18-02-00414, 18-52-00005 and the support of the Ministry of Education and Science of the Russian Federation (GOSZADANIE Grant No. 3.4982.2017/6.7). The development of analytical approach and the calculations of multipole moments have been partially supported by the Russian Science Foundation Grant No. 16-12-10287. Support has been provided by the Government of the Russian Federation (Grant No. 08-08). The research described was performed by Pavel Terekhov as a part of the joint Ph.D. program between the BGU and ITMO.

## Supplementary: The comparison of dynamic and quasistatic multipole sets

It is known that it is possible to describe multipole behavior of dielectric particles either with quasistatic multipole set[1] or dynamic multipole set[2], considering all additions to the multipole moments up to the magnetic quadrupole. Here we compare these approaches for the considered nanocube in air and in medium with the refractive index n = 2. For this purpose, in Fig. S1 we present the scattering cross-sections calculated using both multipole sets and through the numerical far-field integration. In the left side of Fig. S1 one can clearly see that both multipole sets exhibit good coincidence with numerically calculated scattering cross-section. However, when we consider the nanocube in medium with n= 2, the quasistatic approximation much worse coincides with the numerical calculation. In the same time, dynamic multipole approximation provides much better fit, but some discrepancies still take place. This makes it clear that dynamic approximation can be used because of its higher accuracy with respect to the quasistatic approximation; nevertheless, the multipole moments of the higher order provide additional contribution to the light scattering process.

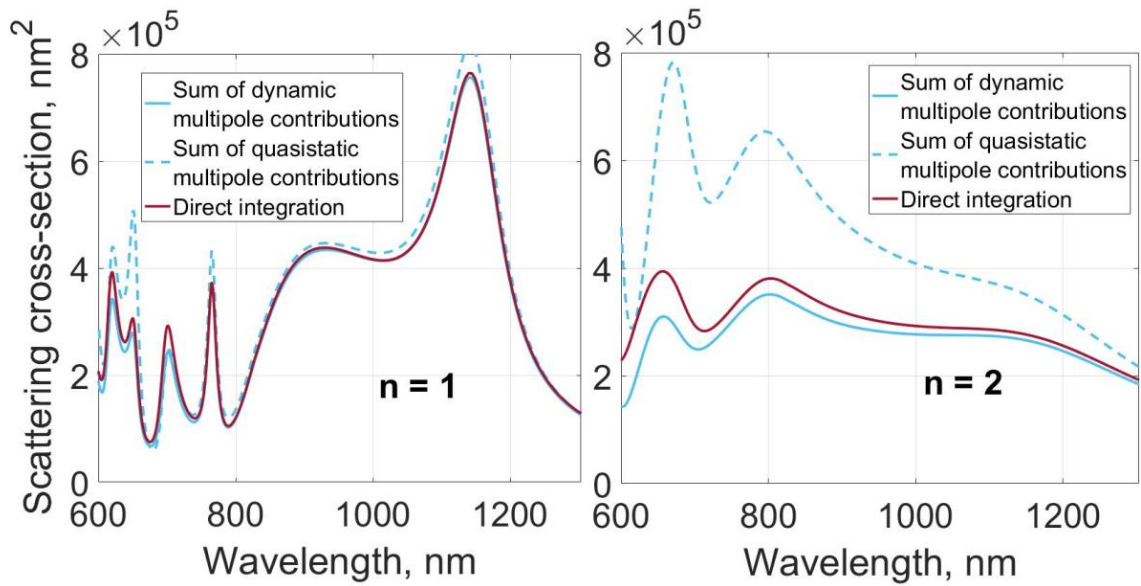

Figure S1: The comparison of the scattering cross-section calculations using both multipole sets and direct integration for n = 1 (left) and n = 2 (right).